\documentclass[pra,twocolumn,superscriptaddress,amsmath,amssymb,showpacs,longbibliography]{revtex4-1}

\usepackage{hyperref}
\pdfoutput=1
\usepackage{graphicx}

\usepackage{tabularx} 
\newcolumntype{Y}{>{\centering\arraybackslash}X} 
\usepackage{colortbl} 

\usepackage{amsmath}
\usepackage{bm}
\usepackage{amssymb}
\usepackage{textcomp}
\usepackage{stmaryrd} 
\usepackage{ulem}
\usepackage{gensymb}
\usepackage{upgreek}

\usepackage{import}
\usepackage{color}
\usepackage{verbatim}

\usepackage{float}

\usepackage{sistyle} 

\begin{document}

\title{Room Temperature Valley Polarization and Coherence in Transition Metal Dichalcogenide-Graphene van der Waals Heterostructures}

\author{Etienne Lorchat}
\altaffiliation{These authors contributed equally to this study}
\affiliation{Universit\'e de Strasbourg, CNRS, IPCMS, UMR 7504, F-67000 Strasbourg, France}

\author{Stefano Azzini}
\altaffiliation{These authors contributed equally to this study}
\affiliation{ISIS \& icFRC, Universit\'e de Strasbourg and CNRS, UMR 7006, F-67000 Strasbourg, France}

\author{Thibault Chervy}
\altaffiliation{These authors contributed equally to this study}
\affiliation{ISIS \& icFRC, Universit\'e de Strasbourg and CNRS, UMR 7006, F-67000 Strasbourg, France}

\author{Takashi Tanigushi}
\affiliation{National Institute for Materials Science, Tsukuba, Ibaraki 305-0044, Japan}

\author{Kenji Watanabe}
\affiliation{National Institute for Materials Science, Tsukuba, Ibaraki 305-0044, Japan}

\author{Thomas W. Ebbesen}
\affiliation{ISIS \& icFRC, Universit\'e de Strasbourg and CNRS, UMR 7006, F-67000 Strasbourg, France}
\author{Cyriaque Genet}
\affiliation{ISIS \& icFRC, Universit\'e de Strasbourg and CNRS, UMR 7006, F-67000 Strasbourg, France}
%
%
\author{St\'ephane Berciaud}
\email{stephane.berciaud@ipcms.unistra.fr}
\affiliation{Universit\'e de Strasbourg, CNRS, IPCMS, UMR 7504, F-67000 Strasbourg, France}

\begin{abstract}
Van der Waals heterostructures made of graphene and transition metal dichalcogenides (TMD) are an emerging platform for opto-electronic, -spintronic and -valleytronic devices that could benefit from (i) strong light-matter interactions and spin-valley locking in TMDs and (ii) exceptional electron and spin transport in graphene. The operation of such devices requires significant valley polarization and valley coherence, ideally up to room temperature.  Here, using a comprehensive Mueller polarimetry analysis, we report \textit{artifact-free} room temperature degrees of valley polarization up to $40~\%$ and, remarkably, of valley coherence up to $20~\%$ in monolayer tungsten disulfide (WS$_2$)/graphene heterostructures. Valley contrasts have been particularly elusive in molybdenum diselenide (MoSe$_2$), even at cryogenic temperatures. Upon interfacing monolayer MoSe$_2$ with graphene, the room temperature degrees of valley polarization and coherence are as high as $14~\%$ and $20~\%$, respectively. Our results are discussed in light of recent reports of highly efficient interlayer coupling and exciton transfer in TMD/graphene heterostructures and hold promise for room temperature chiral light-matter interactions and coherent opto-valleytronic devices.
\end{abstract}

\maketitle


{\textbf{\textit{Introduction} --}} Semiconducting transition metal dichalcogenides (TMD, with formula MX$_2$, where M=Mo,W and X=S, Se, Te) are layered materials endowed with exceptional physical properties, which are promising for innovative two-dimensional opto-electronic and -valleytronic devices~\cite{Mak2016,Schaibley2016}. In particular, monolayer TMD (1L-TMD) exhibit direct optical bandgaps and exciton binding energies around 20 times larger than the room temperature thermal energy~\cite{Wang2018}. Due to the combination of strong spin-orbit coupling and inversion symmetry breaking, 1L-TMD inherit spin-valley locked properties and chiral optical selection rules~\cite{Xu2014}. As a result, valley-polarized excitons~\cite{Mak2012b,Zeng2012b,Cao2012,Sallen2012} and their coherent superpositions~\cite{Jones2013} can be formed using circularly and lineary polarized light, respectively, and further manipulated using external fields~\cite{Ye2016,Wang2016c,Schmidt2016,Hao2016}. 

Unfortunately, in pristine 1L-TMD, valley depolarization and valley decoherence occur on picosecond~\cite{Lagarde2014,Wang2014,Zhu2014b,Molina2017} and sub-picosecond~\cite{Ye2016,Wang2016c,Schmidt2016,Hao2016} timescales, respectively. As a result, robust valley contrasting properties, have chiefly been demonstrated at cryogenic temperatures~\cite{Schaibley2016,Mak2012b,Zeng2012b,Cao2012,Sallen2012,Jones2013,Lagarde2014,Wang2014}, where the exciton lifetime is on the order of a few ps only~\cite{Robert2016}, and where phonon induced intervalley scattering and pure dephasing are minimally efficient. A major challenge is therefore to preserve valley-contrasting properties up to room temperature (RT), where the \textit{effective} exciton lifetime typically exceeds 100~\rm ps in bare 1L-TMD~\cite{Robert2016,Wang2018}.

Room temperature valley polarization has been observed in bare 1L-MoS$_2$~\cite{Sallen2012} or WS$_2$~\cite{Nayak2016,Mccreary2017}, at the cost of a defect-induced reduction of the excitonic lifetime, or, recently, in more complex assemblies, by strongly coupling 1L-WS$_2$ or 1L-MoS$_2$ excitons to an optical mode~\cite{Chervy2018,Chen2017,Sun2017,Lundt2017}. In this case, a cavity protection effect has been invoked to account for RT valley polarization. Noteworthy, valley coherence is directly sensitive to extrinsic and intrinsic pure dephasing mechanisms and hence much more fragile than valley polarization~\cite{Hao2016}. The largest degrees of valley coherence reported to date reach up to 55$\%$ at 4~K~\cite{Cadiz2017}. However, RT  valley coherence has so far eluded experimental observation until our recent report of a steady state degree of valley coherence of $5~\%$ to $8~\%$ in WS$_2$ coupled to a plasmonic array~\cite{Chervy2018}. Overall, obtaining robust RT valley contrasts  in high-quality 1L-TMD is challenging but is, at the same time, a key prerequisite in view of emerging opto-spintronic and -valleytronic devices~\cite{Luo2017,Avsar2017}. Such devices typically interface (i) 1L-TMD as a chiral optical material and/or as an injector of spin/valley polarized electrons with (ii) graphene (Gr), as a high mobility channel for efficient spin-polarized electron transport~\cite{Han2014,Ghiasi2017,Benitez2017}. In view of their obvious relevance for opto-valleytronics, valley polarization~\cite{Du2018} and, crucially, valley coherence in 1L-TMD/Gr heterostructures deserve dedicated investigations.

In this letter, we investigate the valley contrasting properties of van der Waals heterostructures made of 1L-TMD and graphene. In these systems, highly efficient interlayer coupling leads to drastically shortened ($\lesssim 1~\rm ps$) 1L-TMD exciton lifetime~\cite{He2014b,Froehlicher2018} at RT. Valley-polarized excitons can thus quickly recombine radiatively or be directly transferred to graphene before undergoing intervalley scattering and dephasing processes.  Using a comprehensive polarimetry analysis based on the Mueller formalism, we uncover RT degrees of valley polarization up to $40~\%$ and, remarkably, RT degrees of valley coherence up to $20~\%$ in high quality 1L-WS$_2$/Gr heterostructures. Valley contrasts have been particularly elusive in MoSe$_2$, even at cryogenic temperatures~\cite{Wang2015b}. Upon interfacing 1L-MoSe$_2$ with graphene, we observe sizeable RT valley polarization of up to $14~\%$ and valley coherence as high as $20~\%$. Robust RT valley coherence illustrates the high quality and homogeneity of our samples and opens many perspectives for coherent opto-valleytronic devices that take full benefit from the strong light-matter interactions and spin-valley locked properties of TMDs in combination with exceptional electron and spin transport in graphene. 



\begin{figure*}[!ht]
\begin{center}
\includegraphics[width=0.8\linewidth]{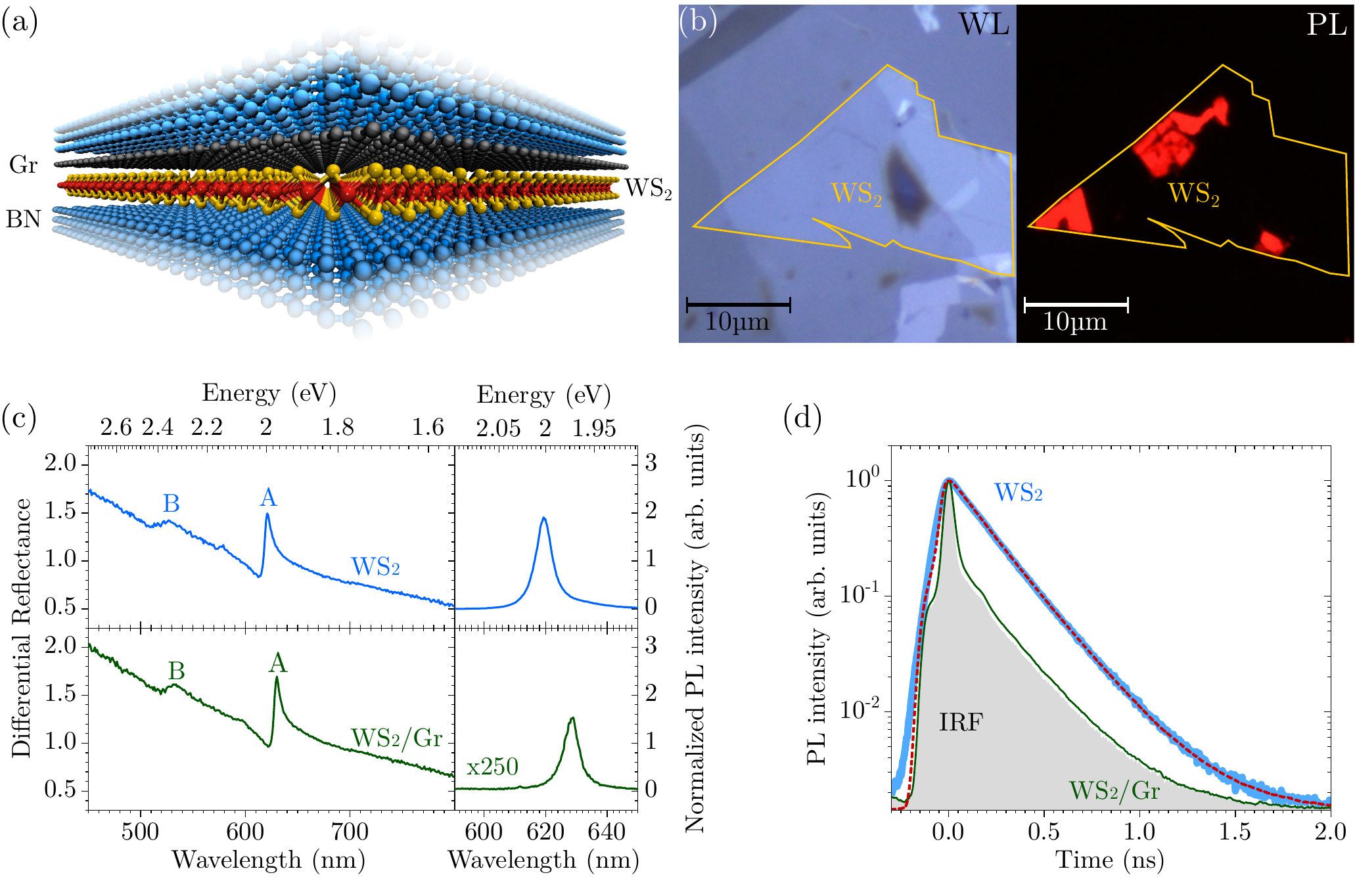}
\caption{(a) Schematic of a BN-capped 1L-WS$_2$/Gr heterostructure.  (b) White light (WL) and photoluminescence (PL) image of a BN-capped 1L-WS$_2$/Gr sample. Dark yellow lines highlight the WS$_2$ monolayer. (c) Differential reflectance (DR) and PL spectra of BN-capped 1L-WS$_2$ (blue) and BN-capped 1L-WS$_2$/Gr (green). The PL spectra were recorded in the linear regime, under \textit{cw} laser illumination at 532~nm (2.33~eV). (d)  PL decay of BN-capped 1L-WS$_2$ (blue solid line) and BN-capped 1L-WS$_2$/Gr (green solid line) recoded under pulsed excitation at 480~nm (2.58~eV). The instrument response function (IRF) is represented by the grey area. The red dashed line is a mono-exponential fit to the BN-capped WS$_2$ PL decay yielding an exciton lifetime of 120~\rm ps.}
\label{Fig1}
\end{center}
\end{figure*}

{\textbf{\textit{Results }--}} 1L-TMD/Gr heterostructures were fabricated from bulk WS$_2$, MoSe$_2$ and graphite crystals using a hot pick-up and transfer method introduced by Zomer \textit{et al.}~\cite{Zomer2014}. In order to get rid of environmental and substrate-induced perturbations, our 1L-TMD/Gr samples were encapsulated using thin layers of hexagonal boron nitride (BN)~\cite{Cadiz2017,Ajayi2017}. The BN/WS$_2$/Gr/BN and The BN/MoSe$_2$/Gr/BN stacks were deposited onto transparent glass substrates so that polarization-resolved photoluminescence (PL) measurements could be performed in a transmission configuration. All measurements described below were performed in ambient air under sufficiently weak incoming photon flux (or pulse fluence) such that non-linear effects such as exciton-exciton annihilation~\cite{Kumar2014} could be neglected.

\begin{figure*}[!thb]
\begin{center}
\includegraphics[width=0.95\linewidth]{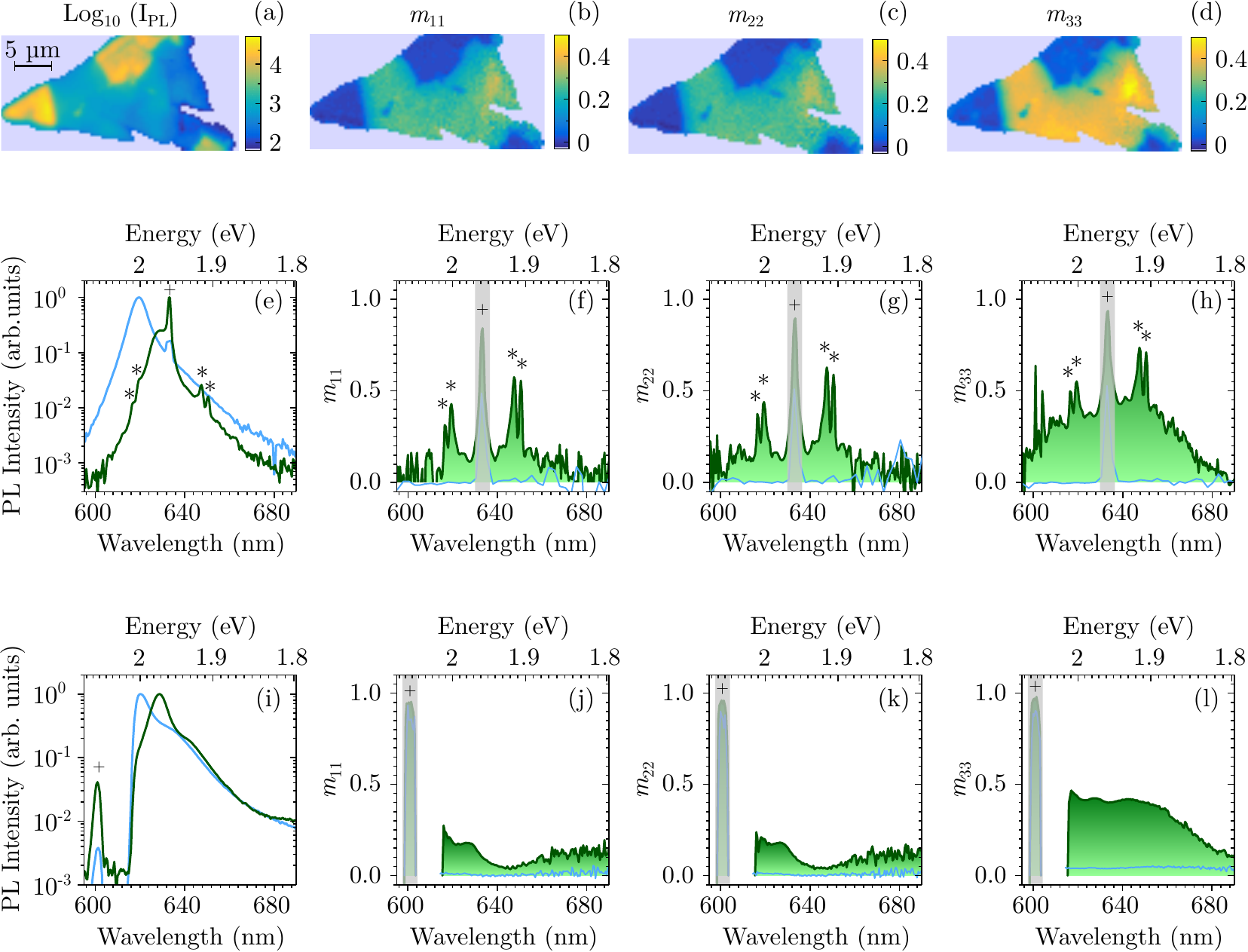}
\caption{Spatially resolved PL (a) intensity  and (b)-(d) diagonal terms of the Mueller matrix ($m_{ii}^{~}$, $i=1,2,3$) of the BN-capped WS$_2$/Gr sample shown in Fig.\ref{Fig1} under optical excitation at 1.96~eV. PL spectra (e) and spectrally-resolved (f)-(h) diagonal terms of the Mueller matrix, under optical excitation at 1.96~eV (e)-(h) and 2.07~eV (i)-(l). The green (blue) curves correspond to BN-capped WS$_2$/Gr (BN-capped WS$_2$). Ultra narrow notch filters (Optigrate), were used for measurements at 1.96~eV (see (e)-(h)) in order to record the full resonant PL spectrum. The $+$ and $*$ symbols  in (e)-(h) highlight residual contributions from the laser beam and polarization contrasts from WS$_2$ Raman scattering features, respectively. Note that, in (j),(k) the slight increase of $m_{11,22}$ on the low-energy wing of the WS$_2$ PL spectrum arises from the faint polarized Raman background from graphene.}
\label{Fig2}
\end{center}
\end{figure*}

Fig.\ref{Fig1} shows (a) the structure and (b) an optical micrograph and a wide-field PL image (obtained using a UV lamp) of the WS$_2$-based sample. Differential reflectance (DR) spectra (recorded using a white light bulb), PL spectra, as well as PL decays are reported in Fig.\ref{Fig1}(c) and (d), respectively. The PL feature arises chiefly from band-edge (A) exciton recombination with a faint redshifted shoulder from charged excitons (trions) (see Fig.~\ref{Fig2}e,i). Due to enhanced dielectric screening from graphene, the PL from BN-capped WS$_2$/Gr (A exciton at 1.98~eV) is slightly redshifted as compared to BN-capped WS$_2$ (A exciton at 2.00~eV)~\cite{Raja2017,Froehlicher2018}. As previously reported, non-radiative exciton transfer from WS$_2$ to graphene leads to massive PL quenching (here, by a factor of $\sim 250$) and reduced exciton lifetime~\cite{He2014b,Froehlicher2018}, well below the temporal resolution of our setup ($\sim~50~\rm ps$). From the $\approx 120~\rm ps$ RT exciton lifetime in BN-capped WS$_2$ we may estimate a RT exciton lifetime  as short as $\sim 500~\rm fs$ in BN-capped WS$_2$/Gr. Similar measurements in BN-capped MoSe$_2$/Gr are reported in the Supporting Information, Fig.~S9.

To date, valley polarization in TMDs has been assessed through measurements of the degree of circular polarization $\rho^{\pm}=\frac{I^{\pm}_{\sigma_\pm}-I^{\pm}_{\sigma_\mp}}{I^{\pm}_{\sigma_+}+I^{\pm}_{\sigma_-}}$, where $I^{pm}_{\sigma_\pm}$, $I^{pm}_{\sigma_\mp}$ denote the polarization-resolved $\sigma_\pm$ components of the total PL intensity, following optical excitation with circularly polarized light ($\sigma_{\pm}$). Similarly, the degree of valley coherence has been considered equal to the degree of linear polarization $\gamma=\frac{I_{\parallel}-I_{\perp}}{I_{\parallel}+I_{\perp}}$, measured under linearly polarized excitation with an arbitrary orientation with respect to the TMD crystal lattice and where $I_{\parallel}$ (resp. $I_{\perp}$) denote the PL intensity for parallel (resp. perpendicular) linear polarizations of the incoming and emitted photons. As explained in the Supporting Information (Sec. S2), this correspondence is only valid in the absence of any contribution from (i) circular or linear  dichroism and (ii) polarization-dependent PL quantum yield. Owing to their highly symmetric hexagonal crystal structure ($D_{3h}$ point group), 1L-TMDs feature isotropic absorption and emission following optical excitation polarized in the layer plane~\cite{Wang2018,Xu2014}. However, polarization artifacts may arise when TMDs are hybridized to related two-dimensional materials, such as graphene.

\begin{figure*}[!thb]
\begin{center}
\includegraphics[width=0.95\linewidth]{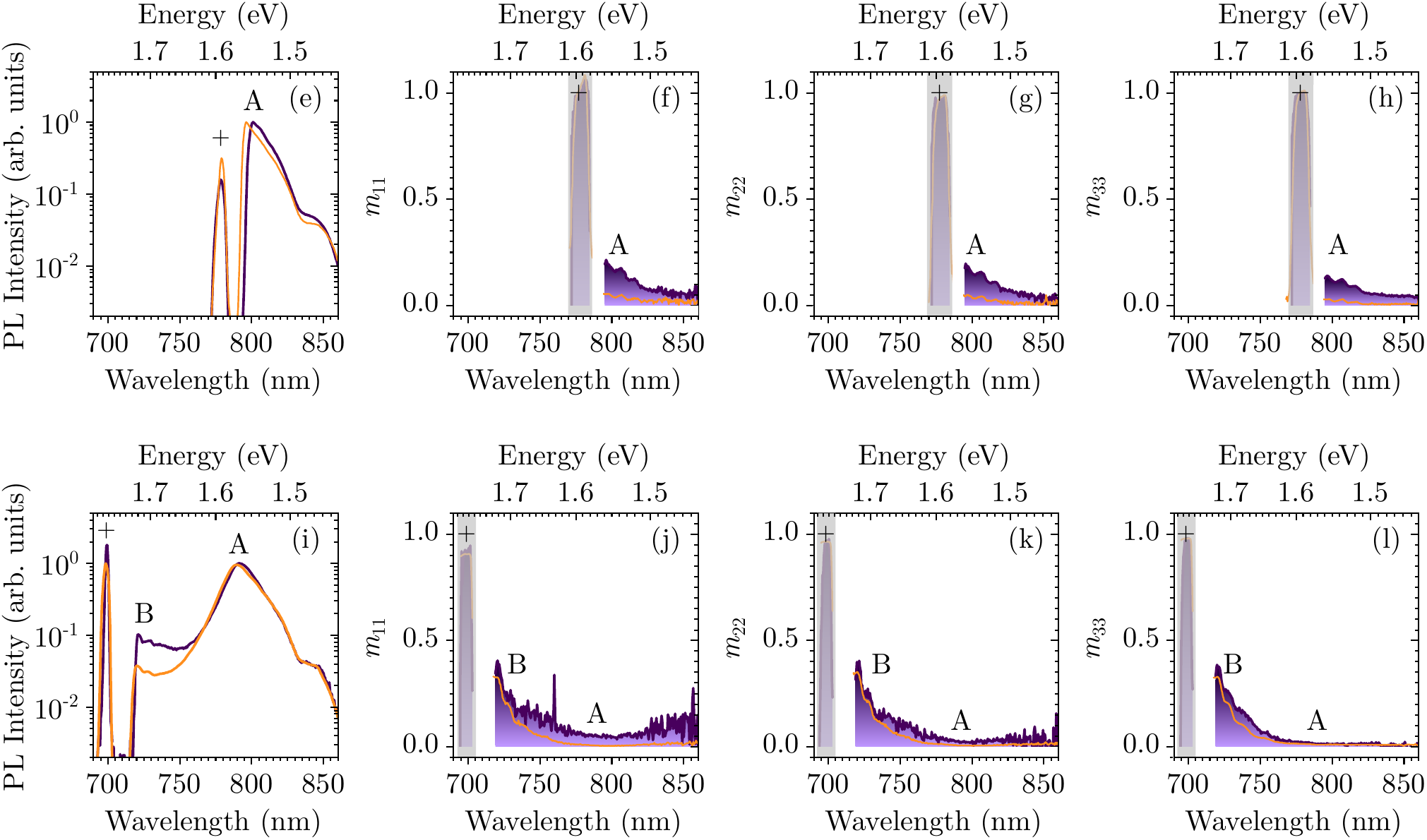}
\caption{PL spectra and  spectrally resolved diagonal terms of the Mueller matrix ($m_{ii}^{~}$, $i=1,2,3$) of a BN-capped MoSe$_2$/Gr sample under optical excitation at 780~nm (i.e., 1.59~eV) (a)-(d), and 700~nm (i.e., 1.77~eV) (e)-(h). The purple (resp. orange) curves correspond to BN-capped MoSe$_2$/Gr (resp. BN-capped MoSe$_2$).The  $+$ symbols highlight residual contributions from the laser beam. The locations of the A and B exciton features are indicated.}
\label{Fig3}
\end{center}
\end{figure*}

In order to provide an \textit{artifact-free} measurement of the valley contrasts, we make use of home-built polarimetry setup that allows us to measure the $4\times4$ Mueller matrix $\mathcal{M}$ associated with the spatially and spectrally resolved  PL response of our samples. The Mueller matrix connects arbitrary incoming polarization states (defined by the Stokes vector of the incoming laser beam) to the outgoing Stokes vector associated with the light emitted by the sample (see Refs.~\citenum{Brosseau1998,LeRoy1997} and Supporting Information, Sec.~S2 and Fig.~S1 for details). In the present study, the most relevant elements of the Mueller matrix are its diagonal terms $m_{ii}^{~}$, with $i=0..3$. By definition, $m_{00}^{~}$ corresponds to the PL intensity and is normalized to unity at all measured wavelengths. Hence, the PL spectra shown in Fig.~\ref{Fig2}a,e and Fig.~\ref{Fig3}a,e in arbitrary units correspond to $m_{00}^{~}$ recorded under unpolarized excitation, without any polarization analysis and prior normalization. With this definition of $m_{00}^{}$, the degrees of valley polarization and valley coherence are directly given by $m_{33}^{~}$ and $m_{11}^{~}$ (or $m_{22}^{~}$), respectively. Circular and linear dichroism are corresponding to $m_{03}^{~}$ and $m_{01}^{~}$, $m_{02}^{~}$, respectively, whereas polarization-dependent PL quantum yields are accounted for by $m_{i0}^{~}, i=1..3$. In the Mueller formalism, $i=1$ (resp. $i=2$) refer to vertical/horizontal (resp. $\pm 45^\circ$) linear polarizations relative to an arbitrary reference angle. Based on symmetry properties, $\mathcal{M}$ is expected to be diagonal in bare 1L-TMDs, with $m_{11}^{~}=m_{22}^{~}$.


Figure~\ref{Fig2} displays the spatially and spectrally resolved diagonal elements of the Mueller matrix of the sample shown in Fig.\ref{Fig1}. The maps in Fig.~\ref{Fig2}a-d correspond to spectrally integrated PL intensity (Fig.~\ref{Fig2}a), valley coherence ($m_{11,22}^{~}$, Fig.~\ref{Fig2}b,c) and valley polarization $m_{33}^{~}$  upon laser excitation at 1.96~eV. A clear anti-correlation between the total PL intensity and the valley contrasts appears, with near zero valley contrasts in BN-capped WS$_2$ (bright regions in Fig.~\ref{Fig1}b and \ref{Fig2}a) and large degrees of valley polarization and coherence in BN-capped WS$_2$/Gr. To quantitatively assess the valley contrasts, we resort to spectrally resolved measurements at two different laser excitation energies, very close to (1.96~eV, i.e., 633~nm, see Fig.~\ref{Fig2}e-h) or slightly above (2.07~eV, i.e., 600~nm, see Fig.~\ref{Fig2}i-l) the optical bandgap of BN-capped WS$_2$/Gr. In stark contrast with the total absence of valley contrasts in our BN-capped WS$_2$ sample ($m_{ii,i=1..3}^{~}\approx0$), BN-capped WS$_2$/Gr exhibits large valley polarization ($m_{33}^{~}\approx 40\%$) \textit{and} coherence ($m_{11}^{~}\approx m_{22}^{~} \approx 20\%$) over the entire span of the PL spectrum. In Fig.~\ref{Fig2}f-h, these contrasts give rise to a baseline on which five sharp peaks with larger contrasts emerge. These  peaks correspond to a faint laser residue, and to polarization sensitive Stokes and anti-Stokes Raman scattering features from (i) the out-of-plane $A'_1$ phonon (near 1.907~eV on the Stokes side) and (ii) the resonant 2LA(M) mode (near 1.915~eV on the Stokes side)~\cite{Berkdemir2013}. Note that the in-plane $E'$ feature expected to overlap with the 2LA(M) feature but has vanishingly small intensity under laser excitation at 1.96 eV. The proposed assignments are unambiguously confirmed by high-resolution polarized Raman measurements (see Supporting Information, Fig.~S8).

Very similar results are observed (Fig.~\ref{Fig2}i-l) when exciting the sample at 2.07~eV, except for the fact that no spurious contributions from Raman features are observed. Similar valley contrasts were observed in another BN-capped WS$_2$/Gr sample and in SiO$_2$-supported WS$_2$/Gr samples either exposed to ambient air (not shown) or covered by a LiF epilayer (see Supporting Information, Fig.~S5).

Motivated by the observation of large RT valley contrasts in WS$_2$/Gr, we now consider the Mueller matrix of MoSe$_2$/Gr. Indeed, no significant valley polarization has been reported so far in bare MoSe$_2$, even at low temperature. The microscopic mechanism responsible for accelerated valley depolarization and decoherence in MoSe$_2$ remains a topic of ongoing research~\cite{Wang2015b,Wang2018}. Figure~\ref{Fig3} shows the PL spectra and $m_{ii,i=1..3}^{~}$ in BN-capped MoSe$_2$/Gr compared to a reference in a BN-capped MoSe$_2$ sample, wherein a short excitonic lifetime (and thus potentially measurable valley contrasts) was observed (see Supporting Information, Fig.~S9). The A exciton in BN-capped MoSe$_2$/Gr (resp. BN-capped MoSe$_2$) is found at 1.568~eV (resp. 1.573 eV) and the higher order B exciton lies near 1.77~eV. Under quasi resonant excitation at 1.59~eV (i.e., 780~nm, see Fig.~\ref{Fig3}a-d), we measure a degree of valley polarization $m_{33}^{~}\approx 14\%$ near the A exciton peak energy in BN-capped MoSe$_2$/Gr. Remarkably, the RT degree of valley coherence $m_{11,22}^{~}\approx 20\%$ in BN-capped MoSe$_2$/Gr exceeds $m_{33}^{~}$. Conversely,  $m_{11,22}^{~}\approx 5\%$ and $m_{33}^{~}\approx 2\%$ in BN-capped MoSe$_2$. 
Under excitation at 1.77~eV (i.e., 700~nm, see Fig.~\ref{Fig3}a-d), slightly above the spin-split B exciton, we observe vanishingly small valley contrasts associated with the A exciton. However, ``hot'' PL from the B exciton exhibits a large degree of valley polarization and coherence, both up to $40~\%$ (resp. $35~\%$) in BN-capped MoSe$_2$/Gr (resp. BN-capped MoSe$_2$). Interestingly, similar valley contrasts are also observed under excitation at 1.50~eV (i.e., 825~nm), slightly below the MoSe$_2$ bandgap (see Supporting Information, Fig.~S7).


\begin{figure}[!htb]
\begin{center}
\includegraphics[width=1.0\linewidth]{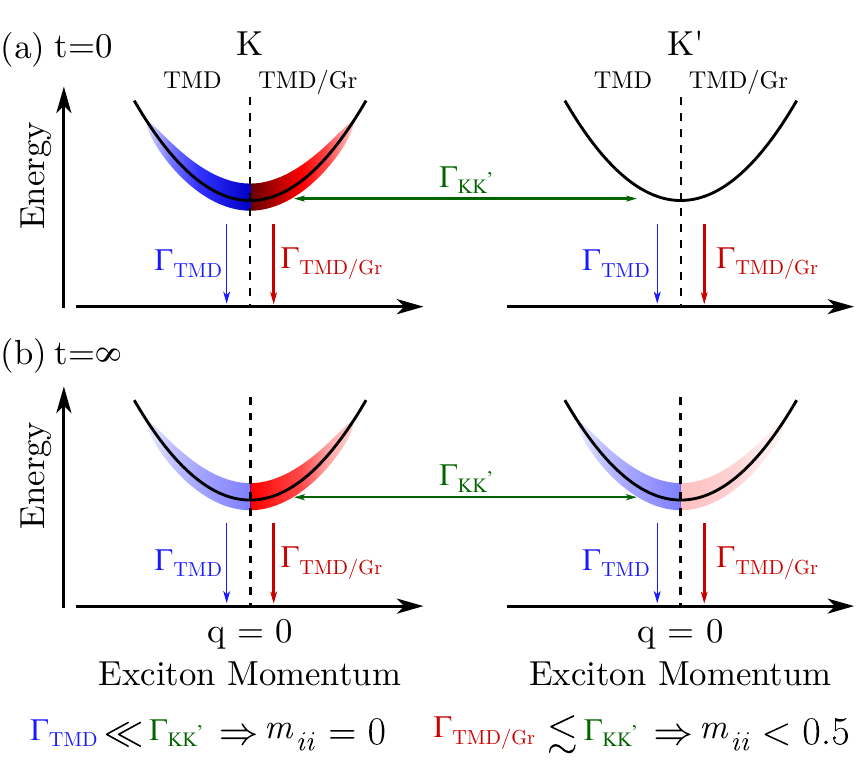}
\caption{Sketches, in the momentum-energy space, of valley-exciton dynamics in bare 1L-TMD compared to 1L-TMD/Gr. $\Gamma_{\rm X}$, with $\rm X=\rm TMD, TMD/Gr$ denote the band-edge exciton decay rate of the bare 1L-TMD and of the 1L-TMD/Gr heterostructure, respectively.  $\Gamma_{\mathrm{\rm KK^{\prime}}}^{~}$ is the intervalley scattering rate. Exciton populations in each valley are illustrated with blue (TMD) and red (TMD/Gr) contours. Darker shades correspond to larger populations. The top panel (a) depicts full valley polarization following optical pumping using circularly polarized photons. The bottom panel (b) represents the populations of valley excitons in the steady state.}
\label{Fig4}
\end{center}
\end{figure}



{\textbf{\textit{Discussion} --}} Our study demonstrates that robust RT valley polarization and coherence can now be generated optically in systems based on 1L-TMD, including in MoSe$_2$, where such contrasts had not yet been reported so far. Importantly, using Mueller polarimetry in 1L-TMD/Gr heterostructures, we experimentally demonstrate that $m_{11}^{~}=m_{22}^{~}$ and that $m_{ij, j\neq i}\approx 0$ (see Supporting Information, Figs.~S2-S4 and~S6), such that circular dichroism and birefringence can be safely neglected in these systems.

At the microscopic level, RT valley contrasts in our TMD/Gr samples are a direct consequence of highly efficient non-radiative exciton transfer from 1L-TMD to Gr~\cite{Froehlicher2018}.  Exciton transfer restricts the observation of radiative recombination of valley polarized excitons and of their coherent superpositions to short ($\lesssim 1~\rm ps$) timescales that are comparable with typical valley polarization and decoherence times. In other words, as illustrated in Fig.~\ref{Fig4}, excitons that would undergo intervalley scattering and dephasing processes in bare 1L-TMD are efficiently filtered out by the near-field coupled graphene layer. Along this line, valley contrasts associated with B exciton emission in MoSe$_2$ and MoSe$_2$/Gr (see Fig.~\ref{Fig3}f-h) also stem from the sub-ps lifetime of these higher-order states. 

Assuming full valley polarization under circularly polarized $cw$ excitation, we can estimate a steady state degree of valley polarization $m_{33}^{~}=\left(1+2\frac{\Gamma_{\rm KK^{\prime}}^{~}}{\Gamma_{\rm X}^{~}}\right)^{-1}$, where $\Gamma_{\rm X}^{~}$ denotes the exciton decay rate and $\Gamma_{\rm KK^{\prime}}^{~}$ the intervalley spin scattering rate, respectively~\cite{Mak2012b} (see also Fig~\ref{Fig4}). Assuming $\Gamma_{\rm X}^{-1}\approx 500~\rm fs$, and considering our measured values of $m_{33}^{~}$, we estimate $\Gamma_{\rm KK^{\prime}}^{-1}\approx 600~\rm fs$ in WS$_2$/Gr and $\Gamma_{\rm KK^{\prime}}^{-1}\approx 150~\rm fs$ in MoSe$_2$/Gr. The fact that the degree of valley coherence $m_{11,22}^{~}$ is approaching (in WS$_2$/Gr) or exceeding (in MoSe$_2$/Gr) $m_{33}^{~}$ indicates that the dominant intervalley scattering mechanism that limits valley polarization is almost exclusively responsible for valley decoherence. Indeed, when pure dephasing is negligible, the valley-exciton decoherence time is twice the lifetime of the valley exciton polarization~\cite{Maialle1993}. Such a near ideal case had so far only been reported in BN-capped MoS$_2$ at 4~K~\cite{Cadiz2017}. At present, intervalley exciton scattering mediated by the exchange interaction is thought to be the dominant valley depolarization and decoherence mechanism~\cite{Zhu2014,Jones2013,Hao2016,Wang2018}. Alternate mechanisms based on electron-phonon interactions have also been proposed~\cite{Molina2017}.  Temperature dependent Mueller polarimetry in TMD/Gr samples should help unravel the relative contributions from both mechanisms.

We note that the valley contrasts reported above come at the cost of significant PL quenching and short exciton lifetimes. Nevertheless, the PL intensity in TMD/Gr has recently been shown to scale linearly with the incident photon fluxes up to values in excess of $10^{24}~\rm cm^{-2}s^{-1}$ (i.e., typically a visible laser beam of 1~mW focused onto a diffraction limited spot~\cite{Froehlicher2018}). In contrast, under these conditions, the PL efficiency of bare 1L-TMD is massively reduced due to exciton-exciton annihilation effects and lies close to that of 1L-TMD/Gr~\cite{Froehlicher2018}. More broadly, 1L-TMD/Gr heterostructures feature major advantages as compared to related systems, in which RT valley contrasts have recently been unveiled. First, even in the absence of  BN capping layers, 1L-TMD/Gr have been shown to be well-defined systems with smooth interfaces and highly reproducible photophysical properties~\cite{Froehlicher2018}. RT valley contrasts can thus consistently be observed in minimally defective TMD/Gr samples. This result is in stark contrast with recent reports in bare 1L-TMD, in which RT valley polarization -and so far, not valley coherence- are observable only when structural defects or extrinsic environmental effects provide sufficiently fast non-radiative exciton decay pathways to bypass intervalley scattering~\cite{Mccreary2017}. Second, RT valley contrasts have also recently been reported in bare bilayer and few-layer TMD~\cite{Zhu2014,Suzuki2014}. In bilayer and even N-layer TMD, inversion symmetry \textit{a priori} precludes the observation of valley contrasting properties and recent observation of circularly polarized emission~\cite{Zhu2014} may either have an extrinsic origin or may be due to A exciton confinement in a 1L unit~\cite{Xu2014} or arise from spin polarization rather than from valley polarization~\cite{Liu2015}. Mueller polarimetry could help settle such debates and determine whether finite degrees of circular or linear polarization in such systems stem from valley polarization and coherence or from the off-diagonal terms of the Mueller matrix. In addition, although RT valley contrasts are symmetry-allowed and have been reported in odd N-layer WS$_2$~\cite{Suzuki2014}, these systems feature indirect optical bandgaps (alike bilayer TMD) and very small PL efficiencies as compared to 1L-TMD/Gr. Last but not least, owing to the excellent electron and spin transport properties of Gr, 1L-TMD/Gr can easily be electrically connected and integrated in chiral light emitting systems~\cite{Zhang2014c} and opto-spintronic circuits~\cite{Luo2017,Avsar2017}.


{\textbf{\textit{Conclusion} --}}
We have demonstrated that monolayer transition metal dichalcogenides (here, WS$_2$ and MoSe$_2$) directly stacked onto monolayer graphene provide highly stable room temperature chiral light emitters, in spite of inevitable photoluminescence quenching. Similar valley contrasting properties are expected in related systems such as MoS$_2$, WSe$_2$ and possibly near-infrared emitters, such as MoTe$_2$~\cite{Ruppert2014,Froehlicher2016}, where, as in bare MoSe$_2$, valley contrasts remain elusive~\cite{Robert2016b}. Our complete analysis, based on the Mueller formalism provides \textit{artifact-free} measurements of valley polarization and valley coherence. As such, it goes beyond state of the art polarimetry in transition metal dichalcogenides, which so far has relied on measurements of circular and linear polarization contrasts~\cite{Neumann2017,Wang2018,Jones2013,Xu2014}. We anticipate further implementations of Mueller polarimetry to investigate chiral light-matter interactions not only in transition metal dichalcogenides (in particular in bilayer or few-layer systems) but also in other emerging two-dimensional systems, such as two-dimensional ferromagnets~\cite{Seyler2018} and van der Waals heterostructures based on the latter~\cite{Zhong2017}. Besides direct implementations in novel opto-valleytronic devices, robust generation of room temperature valley-polarized excitons and, importantly, of valley coherence invite further investigations in nano-photonics, in particular in the chiral strong coupling regime~\cite{Chervy2018}. 

{\textbf{\textit{Supporting Information} --}} Additional details on methods. Mueller polarimetry. Full Mueller matrices measured on BN-capped WS$_2$/Gr and BN-capped MoSe$_2$/Gr. Helicity resolved PL spectra  on WS$_2$/graphene. High resolution Raman measurements on WS$_2$/Gr. Optical characterization of BN-capped MoSe$_2$/graphene.


{\textbf{\textit{Acknowledgements} --}} We thank the STNano clean room staff for technical asistance, J. Hone and D. Chenet for valuable advice on sample fabrication. This work was supported in part by the ANR Equipex ``Union'' (ANR-10-EQPX-52-01), ANR Grant (H2DH ANR-15-CE24-0016), the Labex NIE projects (ANR-11-LABX-0058-NIE) and USIAS within the Investissement d'Avenir program ANR-10-IDEX-0002-02. S.B. is a member of the Institut Universitaire de France (IUF). 



%

\onecolumngrid
\newpage
\begin{center}
{\LARGE\textbf{Supporting Information}}
\end{center}

\setcounter{equation}{0}
\setcounter{figure}{0}
\setcounter{section}{0}
\renewcommand{\theequation}{S\arabic{equation}}
\renewcommand{\thefigure}{S\arabic{figure}}
\renewcommand{\thesection}{S\arabic{section}}
\renewcommand{\thesubsection}{S\arabic{section}.\arabic{subsection}}
\renewcommand{\thesubsubsection}{S\arabic{section}.\arabic{subsection}.\arabic{subsubsection}}
\linespread{1.4}

\bigskip

\section{Methods summary}

We fabricate our BN-capped TMD/Graphene samples by a hot pick-up method introduce by Zomer \textit{et al.}~\cite{Zomer2014}. First, we exfoliate bulk crystals on Silicon wafers coated with a 90~nm oxide epilayer. Then we identify flakes of interest, including TMD and Gr monolayers that we characterize by optical contrast and micro-PL and/or micro-Raman spectroscopy. Starting from the top capping BN layer, layers of choice (graphene, TMD and bottom BN) are sequentially picked up using a polaycarbonate (PC) coated polydimethysiloxane (PDMS) stamp. The stack is finally transferred onto a $170~\mu \rm m$ thick glass coverslip held at a controlled temperature. Finally, PC residues are eliminated in chloroform.

Differential reflectance, PL (including PL decays) and Raman spectra were recorded using a home-built micro-optical spectroscopy setup, as in Ref.~\citenum{Froehlicher2018} . DR measurements were performed using a white-light bulb. PL and Raman spectra were recorded using \textit{cw} lasers either at 532~nm or 633~nm. Time-correlated PL measurements (Fig. 1 and~\ref{FigS9})  were performed using the filtered output of a supercontinuum light source at 480 nm ($\sim$ 50 ps pulse duration) and a single photon counting board. The supercontinuum source was also employed for all polarization-resolved measurements, except those performed at 633~nm, where a \textit{cw} HeNe laser was used. All measurements described below were performed in ambient air under sufficiently weak incoming photon flux (or pulse fluence) such that non-linear effects such as exciton-exciton annihilation could be neglected.

The Mueller polarimetry setup is described in more details below.

	\section{PL Mueller polarimetry}
	
	The optical setup shown in Fig.\ref{FigS1} is used to run Mueller polarimetry of the PL emission from our 1L-MX$_2$/Gr hetero-structures, namely measuring PL spectra for different combinations of excitation and detection polarizations. Such measurements allow us to retrieve the full Mueller matrix  $\mathcal{M}$ of the sample, a spectral characterization of how the polarization state of the emitted PL is related to the
        polarization state of the excitation beam~\citenum{Brosseau1998}, allowing us in particular to retrieve the valley polarization and coherence of the material.
	
	\subsection{Mueller matrix and Stokes vector}
	
	The emission Mueller matrix $\mathcal{M}$ of a given material system determines the polarization state of the PL emission given that a pump beam with a certain polarization state is providing the excitation. An incident excitation in a given polarization state is defined by a Stokes vector $\bf{S}^{\textrm{\tiny in}}$, on which the matrix $\mathcal{M}$ acts to yield an output PL Stokes vector $\bf{S}^{\textrm{\tiny out}}$:
	\begin{equation}
	\bf{S}^{\textrm{\tiny out}} = \left(
	\begin{array}{c}
	I\\
	I_{V} - I_{H}\\
	I_{45} - I_{-45}\\
	I_{\sigma^+} - I_{\sigma^-}
	\end{array}
	\right)_{\textrm{\tiny out}} = \mathcal{M} \cdot \bf{S}^{\textrm{\tiny in}} = \mathcal{M}\left(
	\begin{array}{c}
	I_0\\
	I_{V} - I_{H}\\
	I_{45} - I_{-45}\\
	I_{\sigma^+} - I_{\sigma^-}
	\end{array}
	\right)_{\textrm{\tiny in}}, \label{eq:StokesPL}
	\end{equation}
	where $I_{(0)}$ is the emitted (incident) intensity, $I_{V} - I_{H}$
	is the relative intensity in vertical and horizontal polarizations, $I_{45} - I_{-45}$
	is the relative intensity in $+45^{\circ}$ and  $-45^{\circ}$ polarizations and 
	$I_{\sigma^+} - I_{\sigma^-}$ is the relative intensity in $\sigma^+$
	and  $\sigma^-$ polarizations, and $\mathcal{M}$ is a
        $4\times4$ matrix generally expressed in the form:
	\begin{equation}
	\mathcal{M} = \left(
	\begin{array}{cccc}
	m_{00} & m_{01} & m_{02} & m_{03}\\
	m_{10} & m_{11} & m_{12} & m_{13}\\
	m_{20} & m_{21} & m_{22} & m_{23}\\
	m_{30} & m_{31} & m_{32} & m_{33}
	\end{array}
	\right). \label{eq:Mueller}
	\end{equation}
	
	\subsection{Experimental setup and methods}
	
\begin{figure*}[!tbh]
	\begin{center}
		\includegraphics[width=0.95\linewidth]{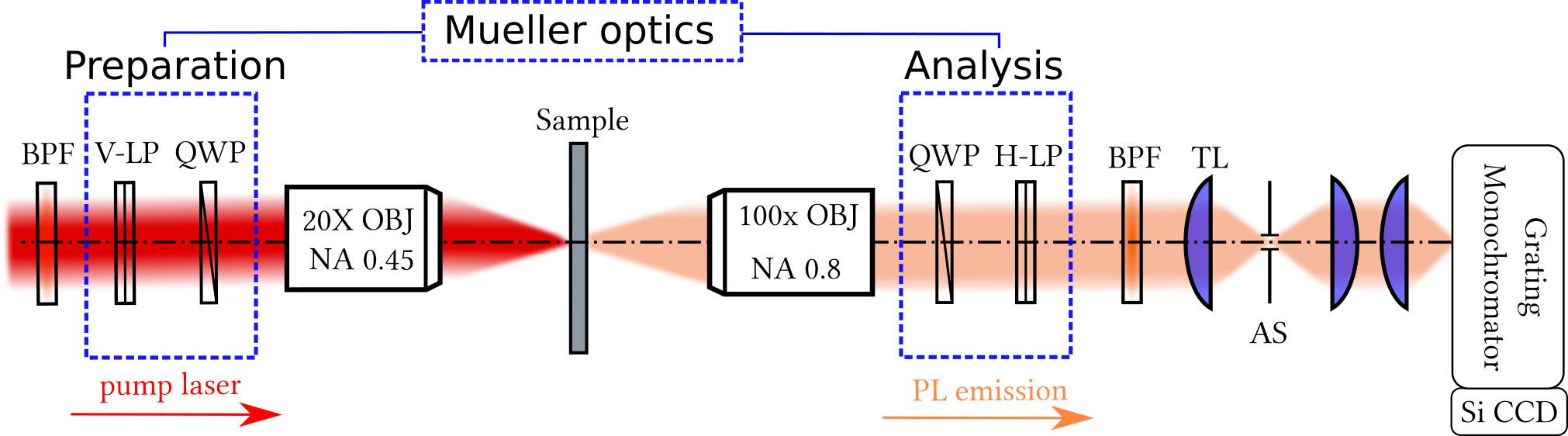}
		\caption{PL Mueller polarimetry setup. BPF: band-pass filter, V-LP/H-LP: vertical/horizontal linear polarizers, QWP: quarter waveplate, TL: tube lens, NA: numerical aperture.}
		\label{FigS1}
	\end{center}
\end{figure*}

	Spectrally resolved Mueller matrices of the PL emission from
        our samples are obtained by means of the optical setup
        sketched in Fig.\ref{FigS1}, where the optical elements
        composing the Mueller polarimeter are highlighted. The
        polarimeter is formed by two stages (both comprising a linear
        polarizer and a quarter-wave plate), one for the preparation
        and another one for the analysis of the polarization state of
        light before and after the sample, respectively. The linear
        polarizer employed in the preparation stage represents the
        first polarization optics on the beam path, which is setting
        the linear state to vertical. On the other hand, in the
        analysis stage, the last element is a linear polarizer set to
        horizontal, ensuring a well-defined polarization state for the
        detection line. Two  broadband quarter-wave plates
        mounted on  rotating
      motors are used for both preparing and 
        analyzing a given polarization state before and after the
        sample, respectively. We adopt a measurement protocol~\cite{LeRoy1997} for
        which each wave-plate is rotated across 8 different angular
        positions of the fast axis by a 22.5\textdegree angular step,
        resulting in a total number of 64 measurements for a full
        reconstruction of the Mueller matrix. For each measurement a
        spectrum is recorded corresponding to the
          $S_{0}$ coefficient of the output Stokes vector for a given
        combination $(\theta_1,\theta_2)$ of
        prepared and analyzed polarization states :
	\begin{equation}
	\bf{S}^{\textrm{\tiny out}} = \mathcal{M}_{\textrm{H-LP}}\mathcal{M}_{\lambda/4}(\theta_2)\mathcal{M}\mathcal{M}_{\lambda/4}(\theta_1)\mathcal{M}_{\textrm{V-LP}}\bf{S}^{\textrm{\tiny in}}.
	\end{equation}
        The Mueller matrices of the horizontal and vertical polarizers
      $(\mathcal{M}_{\textrm{H(V)-LP}})$ are taken as ideal polarizing
    elements (extincition ratio $>10^5:1$), while the wave-plates are
    modelled by homogeneous elliptical birefringent (HEB)
    elements. The wavelength dependent ellipticity and retardance of
    these two HEB waveplates are obtained from transmission
    measurement under white light illumination and in the absence of
    the sample, by minimizing the deviations of the reconstructed Mueller matrix from an identity matrix.
 The algebraic problem is overdetermined,
        consisting of 64 equations for 16 unknowns, so that the
        Mueller matrix is reconstructed by a pseudo-inversion
        operation. The determined parameters of the wave-plates are
        subsequently inserted in the same HEB model which is used for
        getting the Mueller matrix of a PL emission experiment run on
        our 1L-MX$_2$/Gr hetero-structures. Note that
          this broadband calibration proceedure of our Mueller
          polarimeter allows us to accurately account for any
          chromatic distortion of the polarization states
        of both the pump and the PL signal throughout the whole
        calibration range ($550-850$ nm).\\

	\subsection{Valley polarization and coherence from the Mueller matrix}
	
	

The degrees of valley polarization and valley coherence are directly given by $m_{33}^{~}$ and $m_{11}^{~}$ (or $m_{22}^{~}$), respectively. These parameters correspond to the circular ($m_{33}^{~}$) and linear  ($m_{11}^{~}$, $m_{22}^{~}$) polarization contrasts in the absence of any contribution from (i) circular ($m_{03}^{~}$) or linear ($m_{01}^{~}$, $m_{02}^{~}$)  dichroism and (ii) polarization-dependent PL quantum yield (i.e., $m_{i0}^{~}=0$).

	In most reports on valley contrasting properties in TMDs, the valley polarization is assumed to be equal to the degree of circularly polarized PL emission under the same circularly polarized excitation, and it can be expressed using the following formula:

	\begin{equation}
	\rho^\pm = \frac{I_{\sigma^\pm}(\sigma^+) - I_{\sigma^\pm}(\sigma^-)}{I_{\sigma^\pm}(\sigma^+) + I_{\sigma^\pm}(\sigma^-)},
	\label{eq:Circ}
	\end{equation}
	
	where $I_j(l)$ is the measured PL spectrum for a $j=(\sigma^+,\sigma^-)$ polarized excitation and a $l=(\sigma^+,\sigma^-)$ polarized analysis. In a similar way, the valley coherence is usually taken equal to the degree of linearly polarized PL emission under the same linearly polarized excitation, and it can be expressed in two ways according to the chosen pair of linearly polarized states:
	
	\begin{equation}
	\gamma^{VH} = \frac{I_{V,H}(V) - I_{V,H}(H)}{I_{V,H}(V) + I_{V,H}(H)},
	\label{eq:Lin}
	\end{equation}
	\begin{equation}
	\gamma^{+45^{\circ}-45^{\circ}} = \frac{I_{+45^{\circ}-45^{\circ}}(+45^{\circ}) - I_{+45^{\circ}-45^{\circ}}(-45^{\circ})}{I_{+45^{\circ}-45^{\circ}}(+45^{\circ}) + I_{+45^{\circ}-45^{\circ}}(-45^{\circ})},
	\label{eq:Diag}
	\end{equation}
	
	where $I_j(l)$ is the measured PL spectrum for a $j=(V,H)$ or $j=(+45^{\circ},-45^{\circ})$ polarized excitation and a $l=(V,H)$ or $l=(+45^{\circ},-45^{\circ})$ polarized analysis. Eq.~\ref{eq:Lin} is for vertical (V) and horizontal (H) states, while Eq.~\ref{eq:Diag} stands for $+45^{\circ}$ and $-45^{\circ}$ linear polarizations.\\

	Using the Mueller-Stokes formalism, these contrasts can be expressed in terms of the elements of the Mueller matrix, by taking as measured total intensities the first elements $S_{0} = I$ of the outgoing Stokes vectors.

%
	
	Considering that the incoming Stokes vectors for $V$, $H$, $\pm 45^{\circ}$, $\sigma^{\pm}$ polarizations write:
	
	\begin{equation}
	\bf{S}^{\textrm{\tiny in}}_{\textrm{\tiny V,H}} = \left(
	\begin{array}{c}
	1\\
	\pm1\\
	0\\
	0
	\end{array}
	\right)_{\textrm{\tiny V,H}}
	\bf{S}^{\textrm{\tiny in}}_{\textrm{\tiny 45,-45}} = \left(
	\begin{array}{c}
	1\\
	0\\
	\pm1\\
	0
	\end{array}
	\right)_{\textrm{\tiny 45,-45}}
	\bf{S}^{\textrm{\tiny in}}_{\sigma^+,\sigma^-} = \left(
	\begin{array}{c}
	1\\
	0\\
	0\\
	\pm1
	\end{array}
	\right)_{\sigma^+,\sigma^-}, \label{eq:PolInput}
	\end{equation}

	we may then, by fixing the polarization state of the input pump beam, and taking the normalized difference of the total intensities under crossed polarization analysis, we obtain the expressions for $\gamma^{VH}$, $\rho^{45,-45}$ and $\rho^{\pm}$  as a function of the Mueller matrix elements:
	\begin{equation}
	\gamma^{VH} = \frac{m_{10}+m_{11}}{m_{00}+m_{01}},
	\gamma^{+45^{\circ},-45^{\circ}} = \frac{m_{20}+m_{22}}{m_{00}+m_{02}},
	\rho^\pm = \frac{m_{30}\pm m_{33}}{m_{00}\pm m_{03}}.
	\label{eq:contrastsM}
	\end{equation}
	These simple expressions reveal that if the amplitude of the off-diagonal elements of $\mathcal{M}$ is either zero or negligible, the valley polarization and coherence are respectively given by the $m_{33}$ and $m_{11}$ (or alternatively $m_{22}$) elements of the Mueller matrix, normalized with respect to $m_{00}^{~}$. In main the manuscript, we have, at each wavelength, normalized $m_{00}^{~}$ to unity, such that the \textit{unpolarized} PL spectra shown in Fig.~2 and 3 correspond to the \textit{un-normalized} $m_{00}^{~}$.
	
	Owing to their highly symmetric crystal structure, 1L-TMDs feature isotropic absorption and emission following optical excitation polarized in the layer plane~\cite{Wang2018,Xu2014}. As a result, all the off-diagonal elements of $\mathcal{M}$ vanish and $m_{11}^{~}=m_{22}^{~}$. These symmetry arguments justify why, in 1L-TMDs, the degrees of valley polarization ($m_{33}^{~}$) and valley coherence ($m_{11}^{~}$, $m_{22}^{~}$) have thus far been approximated by the helicity parameter $\rho^{\pm}$ and by the degree of linear polarization $\gamma$ (for an arbitrary incoming linear polarization), respectively.
	

\clearpage 
\section{Supplementary Figures}

\begin{figure*}[!tbh]
	\begin{center}
		\includegraphics[width=0.95\linewidth]{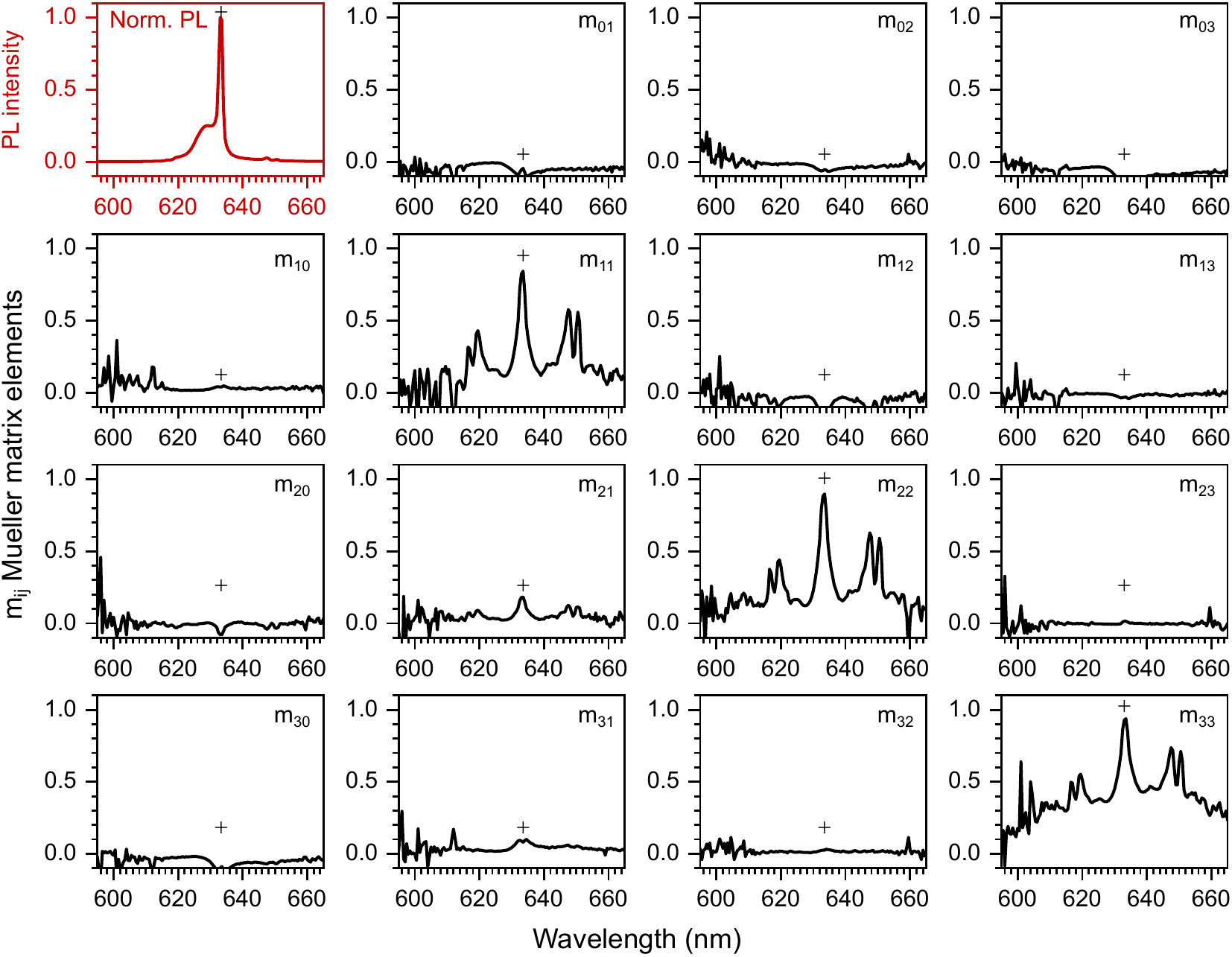}
		\caption{Mueller matrix of the PL emission from our BN-capped WS$_2$/Gr heterostructure recorded in ambient conditions under optical excitation at 633 nm. The $+$ symbols highlight residual contributions from the laser beam.}
		\label{FigS2}
	\end{center}
\end{figure*}

\begin{figure*}[!tbh]
	\begin{center}
		\includegraphics[width=0.75\linewidth]{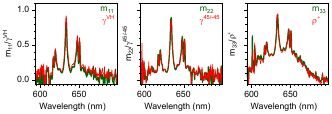}
		\caption{BN-capped WS$_2$/Gr valley polarization and coherence recorded in ambient conditions under optical excitation at 633~nm: comparison between the diagonal terms of the Mueller matrix  (green) and the PL linear and circular polarization contrasts (red) computed according to Eq.~\ref{eq:contrastsM}. The $+$ and $*$ symbols  highlight residual contributions from the laser beam and polarization contrasts from WS$_2$ Raman scattering features, respectively.}
		\label{FigS3}
	\end{center}
\end{figure*}

\begin{figure*}[!tbh]
	\begin{center}
		\includegraphics[width=0.95\linewidth]{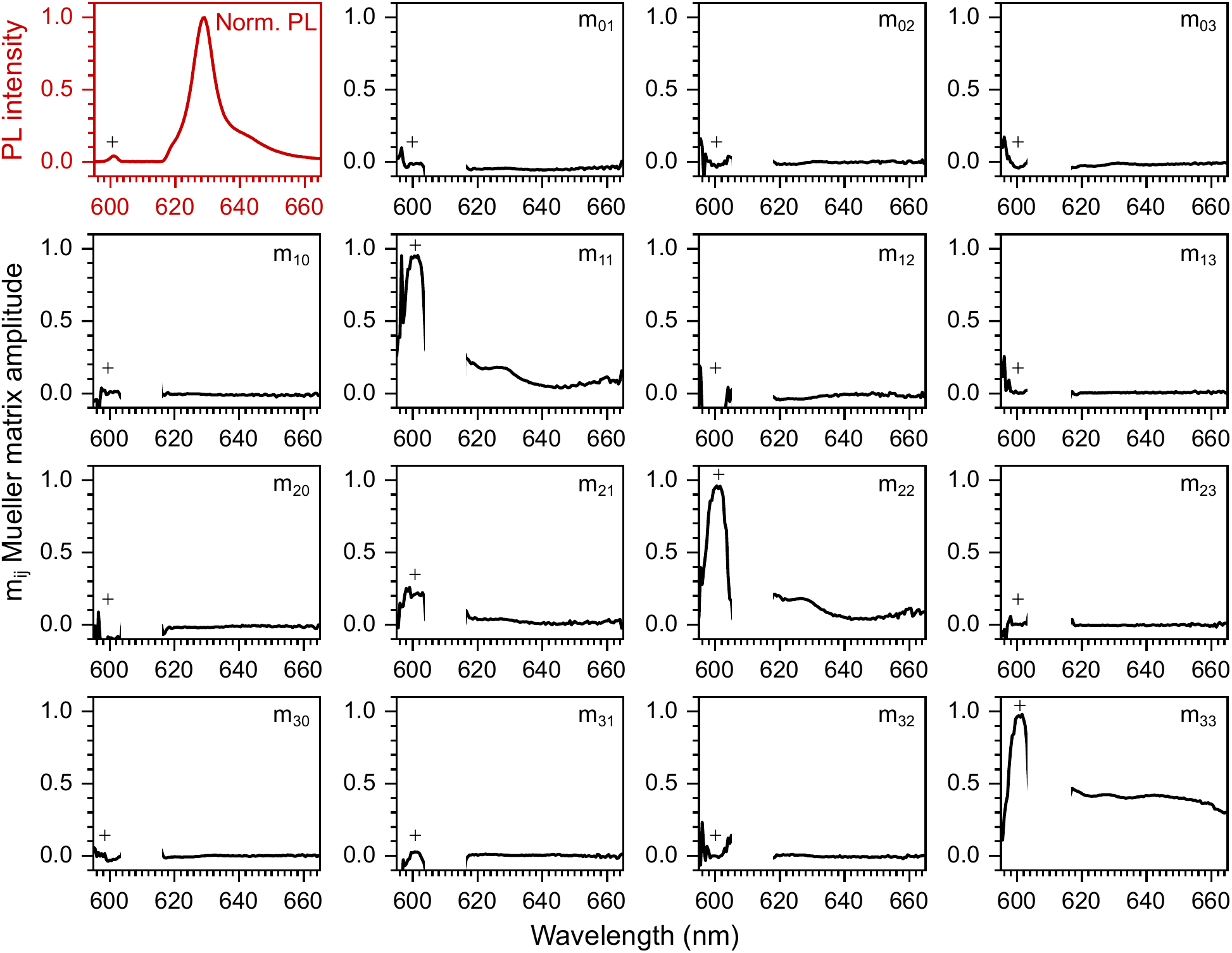}
		\caption{Mueller matrix of the PL emission from our BN-capped WS$_2$/Gr heterostructure recorded in ambient conditions under optical excitation at 600 nm. The  $+$ symbols highlight residual contributions from the laser beam.}
		\label{FigS4}
	\end{center}
\end{figure*}

\begin{figure*}[!tbh]
	\begin{center}
		\includegraphics[width=0.75\linewidth]{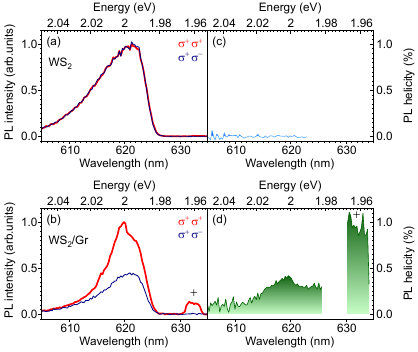}
		\caption{Polarization-resolved PL spectra of (a) WS$_2$ and (b) WS$_2$/Gr recorded in ambient conditions for $\sigma_{\pm}$ polarized light at 1.96~eV (i.e., 633~nm). The samples are excited using $\sigma_{+}$ polarized light. The corresponding degrees of circular polarization are shown in (c) and (d). Significant helicity (up to $25\%$) is only observed in WS$_2$/Gr. The sample is deposited on a SiO$_2$ substrate and capped with a LiF epilayer.}
		\label{FigS5}
	\end{center}
\end{figure*}

	Full Mueller matrices recorded for BN-capped WS$_2$/Gr (see Fig. 2 in the main manuscript) are shown in Fig.~\ref{FigS2} and ~\ref{FigS4}. In Fig.~\ref{FigS3}, we compare, for BN-capped WS$_2$/Gr excited at 633~nm, the computed values of $\gamma^{VH}$, $\gamma^{+45^{\circ},-45^{\circ}}$ and $\rho^+$ are compared to our normalized measurements of $m_{11}$, $m_{22}$, $m_{33}$, respectively. Since the off-diagonal elements of $\mathcal{M}$ are vanishigly small, we indeed do not observe any significant differences between the polarization contrasts and $m_{ii},i=1..3$. In Fig.~\ref{FigS5}, we also show direct measurements of $\rho^+$ for another WS$_2$/Gr sample.

\subsection{Polarimetry measurements on WS$_2$/Gr}

\clearpage

	\subsection{Polarimetry measurements on MoSe$_2$/Gr}
\begin{figure*}[!tbh]
	\begin{center}
		\includegraphics[width=0.95\linewidth]{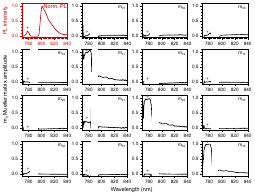}
		\caption{Mueller matrix of the PL emission from our BN-capped MoSe$_2$/Gr heterostructure recorded in ambient conditions under optical excitation at 780 nm. The  $+$ symbols highlight residual contributions from the laser beam.}
		\label{FigS6}
	\end{center}
\end{figure*}

\begin{figure*}[!tbh]
	\begin{center}
		\includegraphics[width=0.95\linewidth]{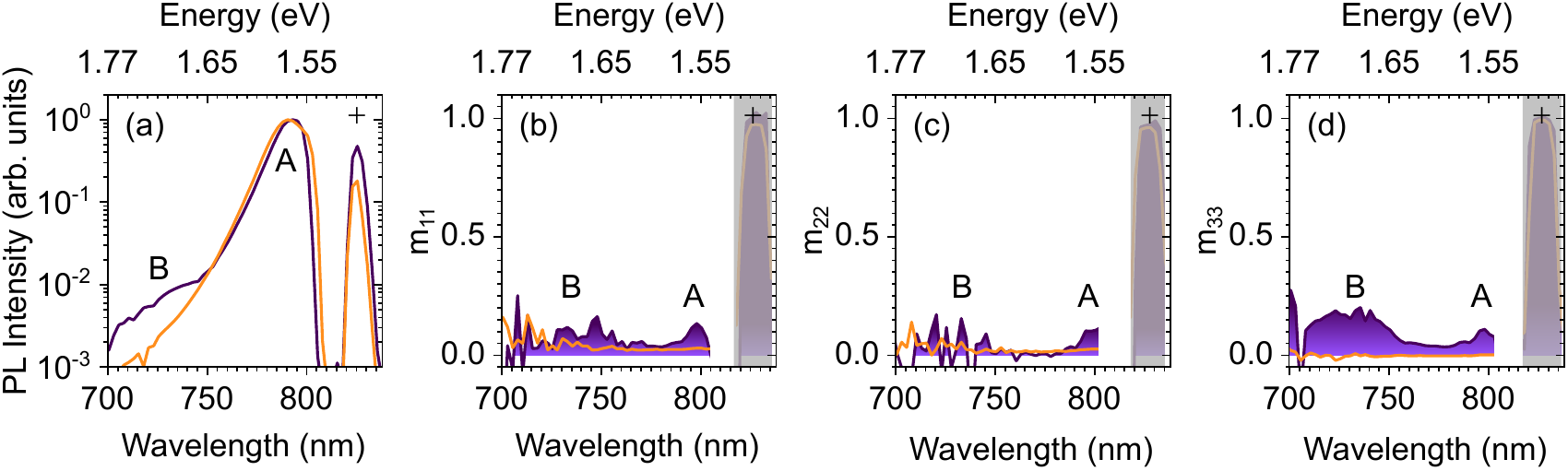}
		\caption{BN-capped MoSe$_2$/Gr valley polarization and coherence: PL spectra ($m_{00}^{~}$) and  spectrally resolved diagonal terms of the Mueller matrix ($m_{ii}^{~}$, $i=1,2,3$) of a BN-capped MoSe$_2$/Gr sample under sub-bandgap optical excitation at 825~nm (i.e., 1.50~eV). The purple (resp. orange) curves correspond to BN-capped MoSe$_2$/Gr (resp. BN-capped MoSe$_2$).The  $+$ symbols highlight residual contributions from the laser beam. Under sub-bandgap excitation, exciton formation may occur through phonon assisted~\cite{Jones2016,Manca2017}  or possibly two-photon upconversion processes~\cite{Manca2017}. We observe nearly equal valley polarization and coherence, both of up to $12\%$ at the A exciton peak energy, in BN-capped MoSe$_2$. Let us note that, unexpectedly, we are still able to resolve very dim upconverted PL from the B exciton in BN-capped MoSe$_2$/Gr, which displays degrees of valley polarization and coherence of up to $16\%$ and $\sim 5-10\%$, respectively. No measurable emission from the B exciton is observed in BN-capped MoSe$_2$ under sub-bandgap excitation.}
		\label{FigS7}
	\end{center}
\end{figure*}

	The Mueller matrix for BN-capped MoSe$_2$/Gr excited at 780~nm is reported in Fig.~\ref{FigS6}. The diagonal terms of the Mueller matrix of MoSe$_2$/Gr excited at 1.50~eV (i.e., 825~nm), slightly below the optical bandgap of MoSe$_2$ are shown in Fig.~\ref{FigS7}.

\clearpage

\subsection{High-resolution Raman scattering spectroscopy on WS$_2$/Gr}

\begin{figure*}[!tbh]
	\begin{center}
		\includegraphics[width=0.8\linewidth]{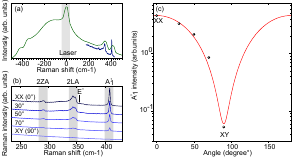}
		\caption{(a) Green: PL spectrum of BN-capped WS$_2$/Gr (see Fig.~1 in the main text). Blue: high-resolution Raman spectrum, recorded in ambient conditions using a spectrometer with 500~mm focal length and a 2400 grooves/mm grating. Both spectra are recorded under linearly polarized optical excitation at 633 nm. The expected position the $E^{\prime}$ mode feature is indicated~\citenum{Berkdemir2013} . (b) Polarization-resolved Raman spectra for parallel (XX) and perpendicular (XY) linear polarizations of the incoming and scattered photons. The main features are indicated, as in Ref.~\citenum{Berkdemir2013}. (c) Integrated intensity of the out-of-plane $A^{\prime}_1$ feature as a function of the angle between the linearly polarized incoming and scattered photons. The solid line is a $\cos^2$ fit to the data.}
		\label{FigS8}
	\end{center}
\end{figure*}

\clearpage
\subsection{Optical characterization of MoSe$_2$/Gr}

\begin{figure*}[!tbh]
	\begin{center}
		\includegraphics[width=0.9\linewidth]{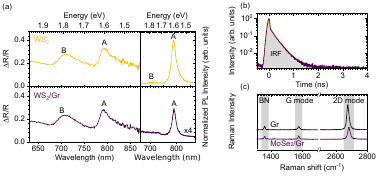}
		\caption{(a) Differential reflectance (DR) and PL spectra of BN-capped 1L-MoSe$_2$ (yellow) and BN-capped 1L-WS$_2$/Gr (purple). The PL spectra were recorded in the linear regime, under \textit{cw} laser illumination at 532~nm (2.33~eV). (b)  PL decay of BN-capped 1L-MoSe$_2$ (yellow) and BN-capped 1L-WS$_2$/Gr (purple) recoded under pulsed excitation at 480~nm (2.58~eV). The instrument response function (IRF) is represented by the grey area. The PL decays of BN-capped MoSe$_2$ cannot be resolved. Nevertheless Significant PL quenching by a factor of ~6 is observed on BN-capped MoSe$_2$/Gr, indicating a shorter exciton lifetime as compared to BN-capped MoSe$_2$/Gr. (c) Raman spectra of BN-capped graphene and BN-capped MoSe$_2$/graphene under optcial excitation at 532 nm.}
		\label{FigS9}
	\end{center}
\end{figure*}

\end{document}